\documentclass[iop]{emulateapj}
\slugcomment{Submitted to ApJ: March 12, 2015}

\usepackage[backref,breaklinks,colorlinks,citecolor=blue]{hyperref}
\usepackage[all]{hypcap}
\usepackage{graphicx} 
\usepackage{mathrsfs} 
\usepackage{natbib}
\usepackage{amsmath}
\usepackage{float}
\usepackage{amssymb}

\newcommand{\lori}{$\lambda$ Orionis}
\newcommand{\hd}{HD 245185}
\newcommand{\OI}{[O\,{\sc i}]}

\newcommand{\as}{$^{\prime\prime}$}

\begin{document}
\title{A SCUBA-2 850-micron Survey of Circumstellar Disks in the $\lambda$ Orionis Cluster}

\author{Megan Ansdell\altaffilmark{1}, Jonathan P.\ Williams\altaffilmark{1}, Lucas A. Cieza\altaffilmark{2}}
\altaffiltext{1}{Institute for Astronomy, University of Hawaii, 2680 Woodlawn Drive, Honolulu, HI 96822, USA}
\altaffiltext{2}{Universidad Diego Portales, Facultad de Ingenier\'{i}a. Av. Ejercito 441, Santiago, Chile}

\email{mansdell@ifa.hawaii.edu}

\begin{abstract}
We present results from an 850-$\mu$m survey of the $\sim$ 5 Myr old \lori{} star-forming region. We used the SCUBA-2 camera on the James Clerk Maxwell Telescope to survey a $\sim0\fdg5$-diameter circular region containing 36 (out of 59) cluster members with infrared excesses indicative of circumstellar disks. We detected only one object at $>3\sigma$ significance, the Herbig Ae star \hd{}, with a flux density of $\sim74$ mJy beam$^{-1}$ corresponding to a dust mass of $\sim150$ M$_{\oplus}$. Stacking the individually undetected sources did not produce a significant mean signal but gives an upper limit on the average dust mass for \lori{} disks of $\sim3$ M$_{\oplus}$. Our follow-up observations of \hd{} with the Submillimeter Array found weak CO 2--1 line emission with an integrated flux of $\sim170$ mJy km s$^{-1}$ but no $^{13}$CO or C$^{18}$O isotopologue emission at 30 mJy km s$^{-1}$ sensitivity, suggesting a gas mass of $\lesssim1$ M$_{\rm Jup}$. The implied gas-to-dust ratio is thus $\gtrsim50$ times lower than the canonical interstellar medium value, setting \hd{} apart from other Herbig Ae disks of similar age, which have been found to be gas rich; as \hd{} also shows signs of accretion, we may be catching it in the final phases of disk clearing. Our study of the \lori{} cluster places quantitative constraints on planet formation timescales, indicating that at $\sim5$ Myr the average disk no longer has sufficient dust and gas to form giant planets and perhaps even super Earths; the bulk material has been mostly dispersed or is locked in pebbles/planetesimals larger than a few mm in size. 

\end{abstract}

\section{INTRODUCTION\label{sec-intro}}

Circumstellar disks are ubiquitous around protostars, being natural consequences of collapsing molecular clouds conserving angular momentum while contracting over two orders of magnitude in size. Space-based infrared (IR) surveys of young stellar clusters have placed strong constraints on disk lifetimes, revealing an exponentially decreasing disk occurrence with age and a characteristic disk lifetime of only a few Myr \citep[e.g.,][]{2014A&A...561A..54R,2009AIPC.1158....3M,2007ApJ...662.1067H}. However, IR observations are sensitive to small amounts of dust due to optical depth effects, saturating at less than a lunar mass of material spread over $\sim100$ AU, making IR excess a poor tracer of total dust mass and thus the planet-forming capacity of disks.

Instead, sub-mm emission is used to probe total disk mass. Continuum emission at these longer wavelengths is optically thin and therefore can be directly related to the total mass of the emitting dust \citep{1983QJRAS..24..267H}. Surveys of young stellar clusters are inconclusive on whether sub-mm emission systematically declines with age, which would reflect steady disk dispersal and grain growth. \cite{2013ApJ...771..129A} showed that, when accounting for different stellar populations and survey sensitivities, the disk populations in Ophiuchus \cite[$\sim1$ Myr;][]{2007ApJ...671.1800A} and IC348 \cite[$\sim2$ Myr;][]{2011ApJ...736..135L} are consistent with that of Taurus \cite[$\sim2$ Myr;][]{2013ApJ...771..129A}, while the discrepancy with the evolved Upper Sco region \cite[$\sim10$ Myr;][]{2012ApJ...745...23M,2014ApJ...787...42C} is only marginal (2.5$\sigma$). However, \cite{2013MNRAS.435.1671W} has also shown that the disk population in the intermediate-aged $\sigma$ Orionis region ($\sim4$ Myr) is distinguishable from that of Taurus at $>3\sigma$ significance. 

It is important to note that estimating total disk mass (gas + dust) from sub-mm continuum flux, which probes only the dust, requires assuming a gas-to-dust ratio. The canonical interstellar medium (ISM) gas-to-dust ratio of $\sim100$ \citep{1978ApJ...224..132B} is often applied, such that $M_{disk}\approx100$ $M_{dust}$. This has been a standard approach as obtaining accurate gas masses is hindered by weak line emission as well as complicated disk chemistry and radiative transfer. Although Class 0/I disks around embedded protostars presumably form with an inherited ISM gas-to-dust ratio, they evolve rapidly to Class II protoplanetary disks around optically visible stars, and eventually to Class III debris disks with negligible gas. Thus this extrapolation of two orders of magnitude is not only large but also uncertain. Using a recently developed technique to independently measure gas masses from CO isotopologues, \cite{2014ApJ...788...59W} found that the gas-to-dust ratios of nine protoplanetary disks in the young ($\sim2$ Myr) Taurus association are already an order of magnitude lower than the ISM value and may vary significantly from disk to disk. This suggests that assuming a constant ISM gas-to-dust ratio, for even very young disks, results in significantly over-estimated total disk masses.

In this paper we focus on the \lori{} cluster in order to further understanding of disk mass evolution and better constrain planet formation timescales. \lori{} is an evolved ($\sim$ 5 Myr; \citealt{2001AJ....121.2124D}) star-forming region, centered on a tight core of OB stars and surrounded by a 30-pc radius ring of dense dust and gas, possibly formed by a supernova that occurred $\sim1$ Myr ago \citep[e.g.,][]{1999AJ....118.2409D}. For our survey we used the Sub-mm Common User Bolometer Array (SCUBA-2) camera on the 15-m James Clerk Maxwell Telescope (JCMT) atop Maunakea, as it combines high sensitivity with large fields of view, enabling unbiased surveys of young stellar clusters on comparable scales to the aforementioned space-based IR surveys. Our study focuses on the area $<0.5^{\circ}$ from the central core, which encompasses the several hundred pre-main sequence (PMS) stars typically associated with the \lori{} cluster. Although the global Initial Mass Function (IMF) of \lori{} is similar to the field, this inner region is marginally overpopulated by OB stars \citep{2001AJ....121.2124D}. The region is too distant for reliable {\it Hipparcos} parallax measurements, thus we adopt the distance of $\sim450$ pc derived from Str\"{o}mgren photometry of the OB stars \citep{2001AJ....121.2124D}. Reddening toward \lori{} is low at $E(B-V)\sim0.12$ \citep{1994ApJS...93..211D} and thus extinction effects should be negligible. See \cite{2008hsf1.book..757M} for a detailed account of the cluster properties. 

We begin by describing our sample selection and observations in Section 2, then present the survey results in Section 3. In Section 4, we compare our results to other young clusters and take a detailed look at our only detection (\hd{}) in order to analyze the implications for disk evolution and planet formation. In Section 5, we summarize our paper and discuss areas of future work.

\section{SAMPLE \& OBSERVATIONS \label{sec-obs}}
 
 \capstartfalse
\begin{deluxetable}{lll}
\tabletypesize{\scriptsize}
\tablecolumns{3}
\tablewidth{220pt}
\tablecaption{Observing log \label{tab-obs}}
\tablehead{
   \colhead{Date} &
   \colhead{$t_{\rm int}$\textsuperscript{a}} &
   \colhead{$\tau_{225}$\textsuperscript{b}} }
\startdata
\cutinhead{JCMT}
2013/09/23  &  1.5  &  0.10 \\
2014/01/17  &  2.2  &  0.20 \\
2014/01/18  &  1.7  &  0.17 \\
2014/01/20  &  0.4  &  0.16 \\
2014/08/23  &  1.3  &  0.09 \\
2014/08/26  &  1.3  &  0.09 \\
2014/10/13  &  2.2  &  0.09 \\
2014/11/12  &  1.8  &  0.13 \\
2015/01/11  &  3.9  &  0.11 \\
2015/04/17  &  1.4  &  0.05 \\
2015/04/19  &  1.8  &  0.04 \\
\cutinhead{SMA}
2015/02/02  &  4.6  &  0.14 
\enddata
\tablenotetext{a}{On-source integration time in hrs.}
\tablenotetext{b}{Average zenith optical depth at 225 GHz.}
\end{deluxetable}
\capstartfalse

 \subsection{Sample selection\label{sec-sample}}

We formed our sample of disk-bearing \lori{} members by combining results from \cite{2009ApJ...707..705H} and \cite{2010ApJ...722.1226H}. They surveyed the region with {\it Spitzer} to identify cluster members and classify their circumstellar disks according to IR excess. \cite{2009ApJ...707..705H} studied the intermediate-mass population, finding 29 members earlier than F5 but only 10 bearing disks. \cite{2010ApJ...722.1226H} studied the lower-mass population, finding 436 members down to the substellar limit but only 49 with disks. This low disk fraction is consistent with the $\sim5$ Myr age of the cluster \citep[e.g.,][]{2014A&A...561A..54R,2009AIPC.1158....3M,2007ApJ...662.1067H}. 

\cite{2009ApJ...707..705H} classified the disks around intermediate-mass members according to their excess emission in {\it Spitzer} IRAC/MIPS bands. They classified nine sources with moderate excess as ``Debris" disks and one source with large excess in all {\it Spitzer} bands as an optically thick ``Primordial" disk. \cite{2010ApJ...722.1226H} classified disks around lower-mass members according to their spectral energy distribution (SED) IR slopes. They classified optically thick disks with the largest excesses as ``Thk," evolved disks with smaller excesses as ``Ev," and (pre-) transition disks with signs of inner disk clearing as ``PTD" or ``TD."  For simplicity, we group these classifications into optically thick disks (Primordial, Thk) and optically thin disks (TD, PTD, Ev, Debris).

\capstartfalse
\begin{figure}
\begin{centering}
\includegraphics[width=8.5cm]{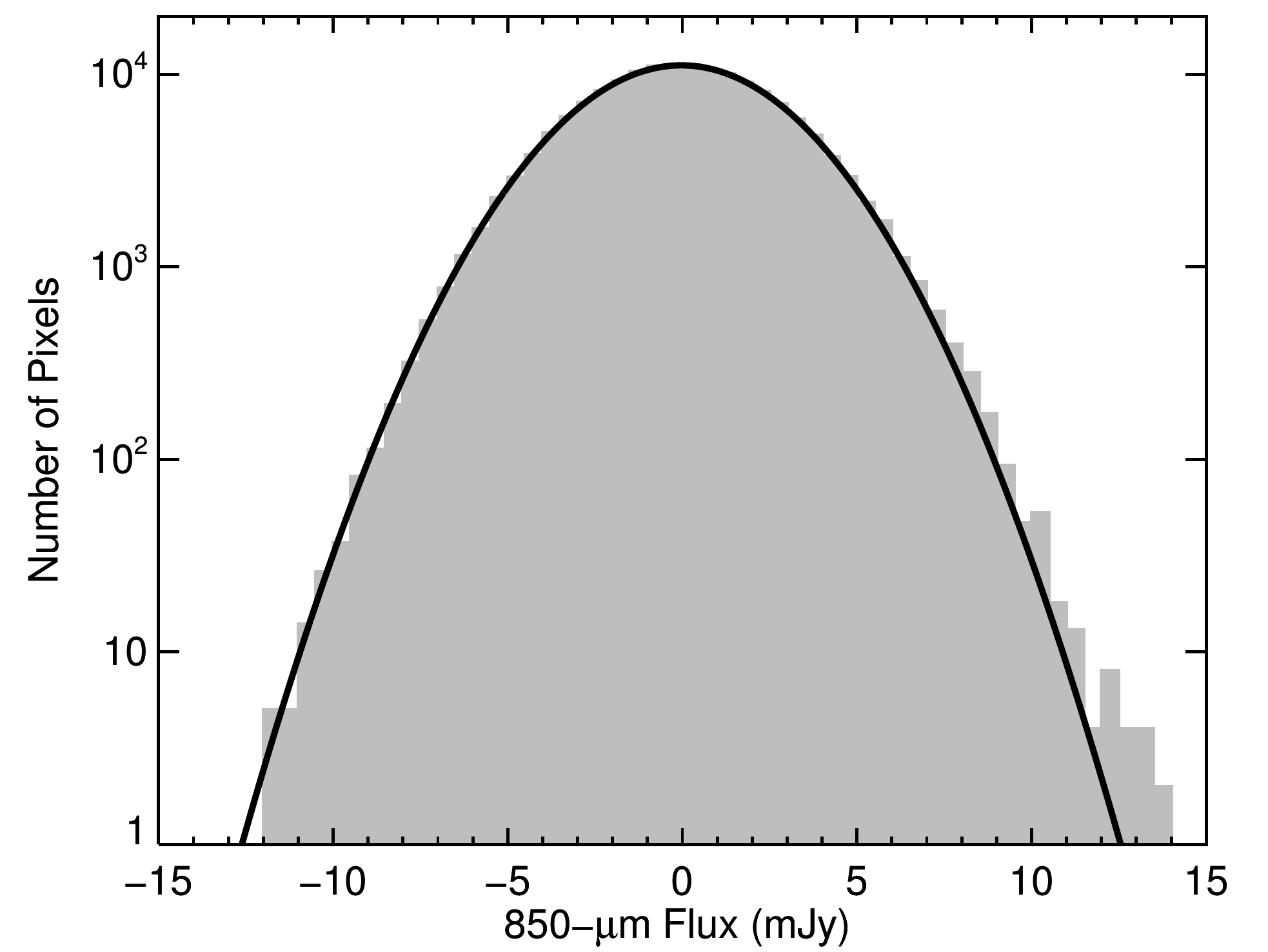}
\caption{\small The histogram of flux densities for all pixels within the usable region of the SCUBA-2 map (Figure~\ref{fig-lori}). The one detection (\hd) was masked out as applying a matched filter (Section~\ref{sec-scuba}) creates negative troughs around strong sources, resulting in a non-Gaussian tail of negative flux densities. The Gaussian fit to the histogram (black line) has a dispersion of 2.9 mJy beam$^{-1}$. The surplus at high flux densities shows the small number of possible detections.\vspace{1.mm}}
\label{fig-hist}
\end{centering}
\end{figure}
\capstartfalse

\capstartfalse
\begin{figure*}
\begin{centering}
\includegraphics[width=17.3cm]{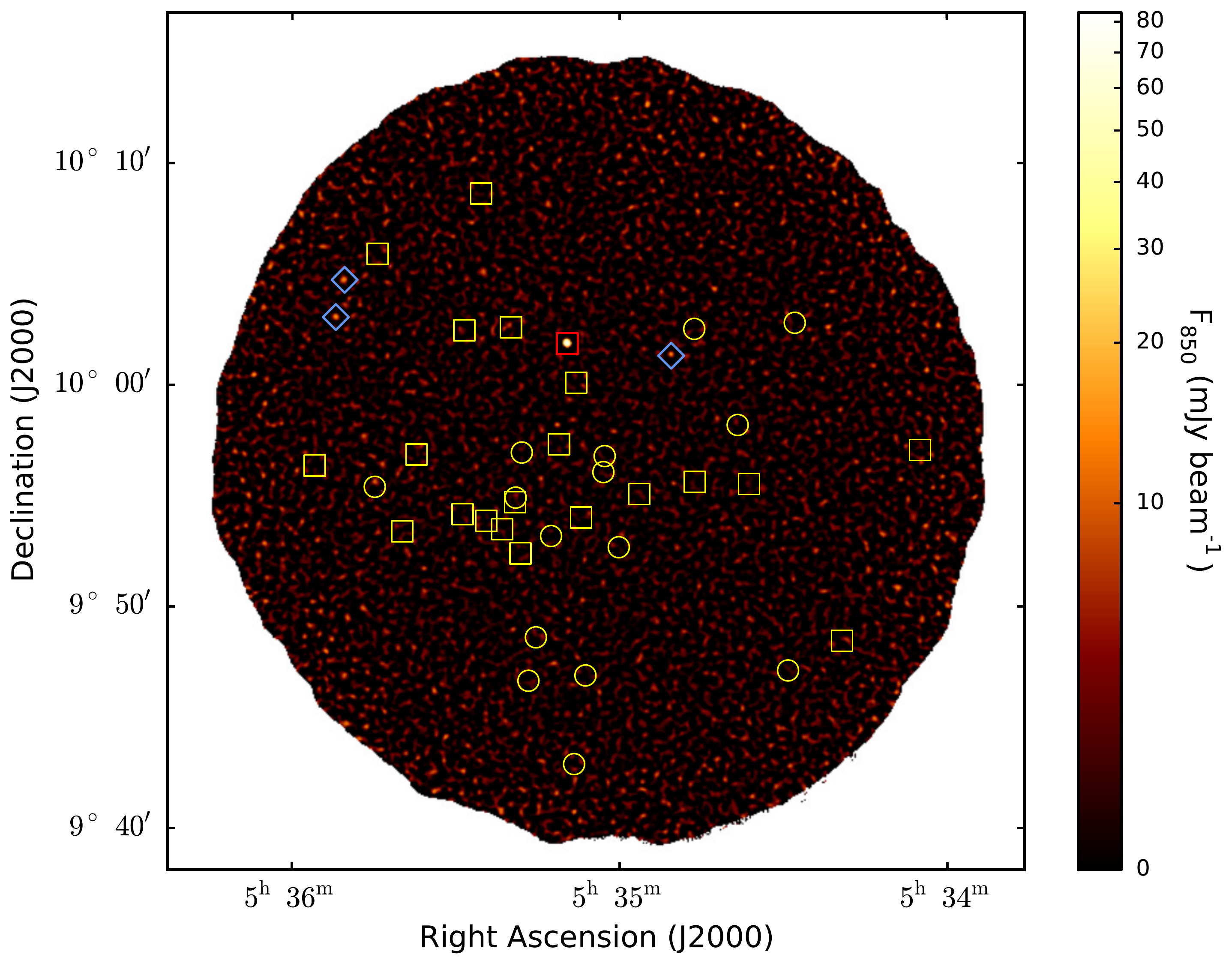}
\caption{\small Our SCUBA-2 850-$\mu$m map of \lori{}, showing locations of disk-bearing members covered by the usable map area (Section~\ref{sec-scuba}). Squares indicate optically thick disks and circles indicate optically thin disks (Section~\ref{sec-sample}). The only significant detection, \hd, is highlighted in red. The blue diamonds indicate possible sub-mm galaxies (Section~\ref{sec-background} \& Table~\ref{tab-background}).}
\label{fig-lori}
\end{centering}
\end{figure*}
\capstartfalse

 \subsection{Observations\label{sec-obs}}

\subsubsection{SCUBA-2\label{sec-scuba}}

We observed a $\sim$ 0\fdg5-diameter circular region toward the \lori{} cluster with SCUBA-2 for a total of $\sim19.2$ hrs. The observations, summarized in Table~\ref{tab-obs}, were taken in queue mode over nine nights from September 2013 to April 2015 (program IDs: M13BH12A, M14AH16A, M14BH050, M15AH01). JCMT defines weather condition bands based on zenith optical depths at 225 GHz ($\tau_{225}$); our data were taken in JCMT band 2 ($0.05<\tau_{225}<0.08$), band 3 ($0.08<\tau_{225}<0.12$), or band 4 ($0.12<\tau_{225}<0.20$) weather. We chose the pointing center of $\alpha$ = 5$^{\rm h}$35$^{\rm m}$4$^{\rm s}$ and $\delta$ = +9$^{\circ}$56$^{\prime}$60$^{\prime\prime}$ to maximize the number of our targets within the map area. The  {\sc pong}1800 mapping mode was used to provide nearly uniform sensitivity over the large circular field. Flats were taken before and after each scan.

We reduced the data using the Dynamic Iterative Map Maker (DIMM) in the {\sc starlink/smurf} software package \citep{2013MNRAS.430.2545C}. The DIMM performs pre-processing and cleaning of the raw data (e.g., dark subtraction, concatenation, flat fielding) to create reduced science maps with a default pixel scale of 4\as{} at 850 $\mu$m. These reduced maps are then flux calibrated and co-added using the Pipeline for Combining and Analyzing Reduced Data (PICARD), a post-processing software also included in the {\sc starlink/smurf} package. 

We reduced each scan individually. We first ran DIMM on the individual chips for a given scan (SCUBA-2 is a mosaic of four detector chips per waveband). We applied a ``blank field" DIMM configuration file, which is commonly used for deep-field extragalactic surveys to enhance faint point source detection (circumstellar disks at the distance of \lori{} should be point sources at the 15\as{} resolution of SCUBA-2 at 850 $\mu$m). We then used PICARD to co-add the reduced chip maps into a single scan map. To maximize point-source detectability, we also applied a matched filter, which suppresses residual large-scale noise by self-subtracting a Gaussian-smoothed version of the map. 

We then flux calibrated each scan. The flux density scale was determined from regular observations of bright point sources, either Uranus or the protoplanetary nebula CRL 618. The calibrators were reduced using the same method described above for the science targets in order to derive a flux calibration factor (FCF), which scales data from units of pW to Jy. Several different FCFs can be applied depending on the data and science needs. Because we applied a matched filter to our science maps, we used FCF$_{\rm match}$, which sets the peak pixel to the total flux density of the point source. FCF$_{\rm match}$ is very stable against beam distortion; we found FCF$_{\rm match}=672\pm36$ Jy pW$^{-1}$ beam$^{-1}$ across our observations, and this $\sim$ 5\% variation represents the absolute flux calibration uncertainty. We applied an FCF$_{\rm match}$ to each reduced scan map, using the calibrator observed closest in time. Before applying the calibration, we inspected the integrity of each calibrator by re-reducing them using the ``bright compact" DIMM configuration file and checking that the resulting FCF$_{\rm peak}$ values matched the nominal values from \cite{2013MNRAS.430.2534D} to within 5\%. 

After each scan was reduced and flux calibrated, we used PICARD to combine all the scans for a given night, then combined all the nightly maps into our final 850-$\mu$m map, shown in Figure~\ref{fig-lori}.  The usable map area was determined by iteratively cropping the noisy outer map regions until the histogram of pixel flux densities represented a Gaussian distribution (Figure~\ref{fig-hist}). This resulted in a $\sim$ 0\fdg45-diameter map with a median rms noise of $\sim2.9$ mJy beam$^{-1}$. The map contains 36 of the 59 known \lori{} members with circumstellar disks (Section~\ref{sec-sample}). These sources are listed in Table~\ref{tab-flux}.

Although SCUBA-2 simultaneously images both 850- and 450-$\mu$m wavebands, the weather conditions were too poor for a 450-$\mu$m survey. However, we recovered our one 850-$\mu$m detection (\hd{}; Section~\ref{sec-det}) at this shorter wavelength and thus reduced the 450-$\mu$m map in a similar manner as for the 850-$\mu$m map described above.

\subsubsection{Submillimeter Array\label{sec-sma}}

We followed up our only SCUBA-2 detection (\hd{}; Section~\ref{sec-det}) with interferometric observations using the Submillimeter Array (SMA). We used the SMA 1300-$\mu$m band with the correlator configured to also cover CO 2--1 line emission at 230.538 GHz. This allowed us to measure the continuum flux at a slightly longer wavelength to extend the SED (Figure~\ref{fig-sed}) as well as search for molecular gas in the disk. 

Data were taken on 2015-02-02 in extended configuration with baselines of $\sim$ 30--200 m. Weather conditions were average with precipitable water vapor of $\sim 2$ mm. The observational setup was typical for SMA, consisting of a long integration on a passband calibrator (3C279) followed by 20-min integrations on the source interleaved with 3--5-min integrations on two gain calibrators (0510+180 and 3C120). The absolute flux scale was derived from observations of Ganymede and is accurate to $\sim20$\%. The data reduction followed standard procedures, using the MIR software package to calibrate the visibilities and MIRIAD for final imaging. The beam sizes of the final maps were $1\farcs00\times0\farcs86$ for the 1300-$\mu$m continuum and $0\farcs90\times0\farcs82$ for the CO 2--1 line emission.

\section{Results \label{sec-results}}

\subsection{SCUBA-2 \label{sec-scubaresults}}

For each source, we measured the total flux density using the pixel value at its Two Micron All-Sky Survey \citep[2MASS;][]{2006AJ....131.1163S} source location. We took this approach because we calibrated our science maps using FCF$_{match}$ (Section~\ref{sec-scuba}), which sets the peak pixel value to the total flux density for point sources. We took the value at the 2MASS source location, rather than searching for the peak pixel within the beam, as the SCUBA-2 850-$\mu$m pixel size (4\as{}) is larger than the combined 2MASS positional errors ($<1$\as{}) and JCMT nominal pointing performance (1.5--2.0\as{}). We estimated 1$\sigma$ statistical uncertainties from the rms noise within an annulus between 20\as{} and 60\as{} in radius. 

The results are tabulated in Table~\ref{tab-flux}. Only one source, the Herbig Ae star \hd, was detected with $>3\sigma$ significance (F$_{850}=74.07\pm2.63$ mJy). We were also able to detected \hd{} at 450 $\mu$m to $\sim6\sigma$ significance (F$_{450}=326\pm52$ mJy). These are the first sub-mm observations of \hd{}, although the target is a known mm source (Section~\ref{sec-smaresults}).

We stacked the individually non-detected sources on their 2MASS positions, then measured the stacked flux ($F_{stack}$) and associated statistical error ($\sigma_{stack}$) in the same manner as described above for individual sources in order to check for a statistically significant mean signal. As reported in Table~\ref{tab-stack}, we did not find a significant mean signal in the stacked image, even when considering only non-detections with optically thick disks. We verified this by calculating the mean flux density ($F_{\mu}$) and standard error on the mean ($\sigma_{\mu}$) for all the non-detections, as well as just the non-detections with optically thick disks, the results of which are also listed in Table~\ref{tab-stack}.

 \capstartfalse                                                           
\begin{deluxetable}{lrrlrrl}                                              
\tabletypesize{\footnotesize}                                           
\centering                                                               
\tablewidth{240pt}                                                         
\tablecaption{SCUBA-2 850-$\mu$m Survey Results \label{tab-flux}}                  
\tablecolumns{7}                                                         
\tablehead{                                                              
   \colhead{ID\textsuperscript{a}}                                      
 & \colhead{RA\textsuperscript{b}}                                                          
 & \colhead{Dec\textsuperscript{b}}                                                         
 & \colhead{SpT\textsuperscript{c}}                                                   
 & \colhead{F$_{850}$\textsuperscript{d}}                                                   
 & \colhead{$\sigma_{850}$\textsuperscript{d}}                                              
 & \colhead{Class\textsuperscript{a}} }                                                           
\startdata                                                               
1624     & 83.5207 & 9.9511  & M7.0 & 4.57 & 3.39 & Thk \\       
2088     & 83.5803 & 9.8076  & M7.0 & 3.92  & 2.71 & Thk \\        
2357     & 83.6163 & 10.0467 & ...  & -2.44 & 2.93 & Ev \\         
2404     & 83.6215 & 9.7853  & ...  & 3.00  & 2.83 & Ev \\         
2712     & 83.6597 & 9.9699  & M5.0 & 2.70  & 2.50 & Ev \\         
2989     & 83.6926 & 9.9271  & M5.5 & -2.99 & 2.40 & Thk \\        
2993     & 83.6930 & 10.0422 & M5.0 & -0.11 & 2.88 & Ev \\         
3360     & 83.7350 & 9.9179  & M4.0 & 4.77  & 2.44 & Thk \\        
3506     & 83.7507 & 9.8780  & ...  & -0.96 & 2.83 & Ev \\         
3597     & 83.7614 & 9.9466  & M4.0 & -2.09 & 2.42 & Ev \\         
3710     & 83.7762 & 9.7816  & M5.5 & -1.80 & 2.91 & Ev \\         
3746     & 83.7795 & 9.9004  & ...  & -0.62 & 2.55 & Thk \\        
3785     & 83.7847 & 9.7149  & ...  & -6.35 & 3.43 & Ev \\         
3887     & 83.7964 & 9.9554  & M5.0 & -1.87 & 2.95 & Thk \\        
3942     & 83.8023 & 9.8864  & M3.0 & -2.95 & 2.35 & Ev \\         
4021     & 83.8139 & 9.8103  & M3.0 & 2.90  & 2.73 & PTD \\        
4111     & 83.8247 & 9.9492  & ...  & 3.35  & 2.52 & TD \\         
4126     & 83.8258 & 9.8734  & M2.0 & -2.89 & 2.13 & Thk \\        
4155     & 83.8294 & 9.9153  & K2   & 6.11  & 2.63 & Ev \\         
4163     & 83.8298 & 9.9118  & M4.0 & 2.17  & 2.54 & Thk \\        
4187     & 83.8330 & 10.0435 & M1.0 & 5.17  & 3.43 & Thk \\        
4255     & 83.8397 & 9.8914  & ...  & -3.74 & 2.31 & Thk \\        
4363     & 83.8517 & 9.8977  & M5.5 & -1.25 & 2.79 & Thk \\        
4407     & 83.8557 & 10.1440 & ...  & -1.73 & 2.95 & Thk \\        
4520     & 83.8686 & 10.0410 & M3.5 & 2.48  & 2.83 & Thk \\        
4531     & 83.8699 & 9.9028  & M5.5 & 1.11  & 2.64 & Thk \\        
4817     & 83.9051 & 9.9477  & ...  & 3.55  & 2.89 & Thk \\        
4916     & 83.9160 & 9.8900  & M6.5 & -0.67 & 2.72 & Thk \\        
5042     & 83.9348 & 10.0984 & M6.0 & -0.41 & 2.96 & Thk \\        
5447     & 83.9827 & 9.9394  & M3.5 & -3.56 & 3.09 & Thk \\        
7951     & 83.7831 & 10.0017 & M6.0 & -1.14 & 2.58 & Thk \\        
7957     & 83.6512 & 9.9256  & M8.0 & 2.31  & 2.72 & Thk \\        
HD36894  & 83.8196 & 9.7776  & B2   & -0.58 & 2.90 & Debris \\     
HD245168 & 83.7624 & 9.9345  & B9   & -1.36 & 2.47 & Debris \\     
HD245185 & 83.7900 & 10.0310 & A0   & 74.07 & 2.63 & Primordial \\ 
HD245275 & 83.9369 & 9.9234  & A5   & 5.27  & 2.72 & Debris
\enddata      
\tablenotetext{a}{From \cite{2009ApJ...707..705H} and \cite{2010ApJ...722.1226H}.}                                     
\tablenotetext{b}{Units of deg (J2000).}                                     
\tablenotetext{c}{Spectral types from literature \citep{2008A&A...488..167S,2007ApJ...664..481B,2011A&A...536A..63B,2011A&A...530A.150F}}                                                           
\tablenotetext{d}{Units of mJy beam$^{-1}$.}                                           
\end{deluxetable}                                                        
 \capstartfalse

\capstartfalse
\begin{deluxetable}{llllll}
\tablecolumns{6}
\tablewidth{245pt}
\tablecaption{Stacking analysis \label{tab-stack}}
\tablehead{
   \colhead{Type}
   & \colhead{N\textsuperscript{a}}
   & \colhead{$F_{stack}$\textsuperscript{b}}
   & \colhead{$\sigma_{stack}$\textsuperscript{b}}
   & \colhead{$F_{\mu}$\textsuperscript{b}}
   & \colhead{$\sigma_{\mu}$\textsuperscript{b}} }
\startdata
Thick  &  19  &  0.46  &  0.72 & 0.18 & 0.60  \\
All      &   35  &  0.40  &  0.46 & 0.30 & 0.46
\enddata
\tablenotetext{a}{Number of non-detected sources used.}                                     
\tablenotetext{b}{Units of mJy beam$^{-1}$. \vspace{1mm}}                                     
\end{deluxetable}
\capstartfalse
           
\subsection{Submillimeter Array\label{sec-smaresults}}
 
\begin{figure}
\begin{centering}
\includegraphics[width=8.cm]{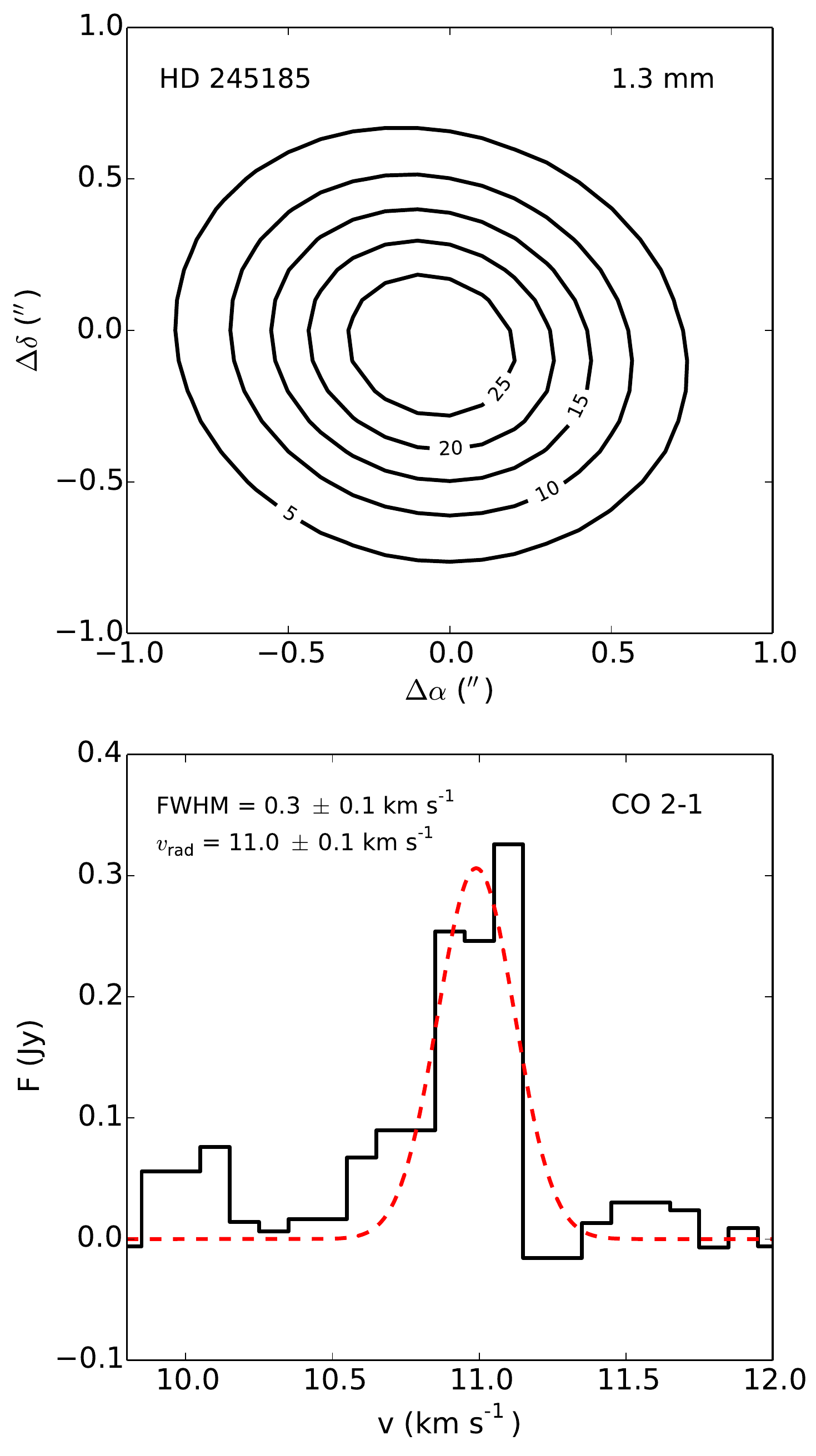}
\caption{\small Results from our SMA observations of \hd{} (Section~\ref{sec-smaresults}). The top panel shows the strong but unresolved 1300-$\mu$m continuum emission where ${F_{1300}}=30\pm6$ mJy (contours in units of mJy). The bottom panel shows the CO 2--1 line emission at 230.538 GHz with a weak integrated flux of $170\pm30$ mJy km s$^{-1}$. A fitted Gaussian (dashed red curve) is centered at $11.0\pm0.1$ km s$^{-1}$ with a narrow velocity profile of ${\rm FWHM}=0.3\pm0.1$ km s$^{-1}$.\\}
\label{fig-sma}
\end{centering}
\end{figure}

Figure~\ref{fig-sma} shows the results from our SMA observations. We clearly detected 1300-$\mu$m continuum emission from \hd{} where ${F_{1300}}=30\pm6$ mJy. \hd{} has been previously detected at mm wavelengths: \cite{1997ApJ...490..792M} reported ${F_{1300}}=44\pm12$ mJy and ${F_{2600}}=6.5\pm1.2$, which is consistent with our results. The source was unresolved in the $1\farcs00 \times 0\farcs86$ SMA beam, thereby constraining the disk radius to $\lesssim200$ AU, which is an improvement on the previous upper limit of $\sim680$ AU from \cite{1997ApJ...490..792M}. 

Weak CO 2--1 line emission was found at $v_{\rm rad}=11.0\pm0.1$ km s$^{-1}$ with a peak of $0.31\pm0.05$ Jy and a velocity resolution of 0.1 km s$^{-1}$. The velocity profile is very narrow with ${\rm FWHM}=0.3\pm0.1$ km s$^{-1}$, suggesting a face-on disk geometry. The integrated line flux is weak at $170\pm30$ mJy km s$^{-1}$. No $^{13}$CO or C$^{18}$O emission was detected at the rms noise level of 30 mJy km s$^{-1}$.

\capstartfalse
\capstartfalse

\subsection{Background sources \label{sec-background}}

Sensitive, large-scale SCUBA-2 surveys similar to this work are used to search for faint sub-mm galaxies (SMGs). These dusty, star-forming galaxies at high redshift are interesting because they are thought to have been instrumental in the build-up of massive galaxies \citep{2014PhR...541...45C}. Based on our SCUBA-2 map size, we expected one 4$\sigma$ noise spike \citep{2013ApJ...762...81C} and thus used a 4.5$\sigma$ threshold for detection of possible SMGs. We found 3 candidate SMGs that lacked counterparts within 15\as{} at all other wavelengths. This matches expectations from the number counts model of \cite{2013ApJ...776..131C} and is the same number found by \cite{2013MNRAS.435.1671W} in their SCUBA-2 map of $\sigma$ Orionis with similar size and sensitivity. The locations and fluxes of these candidate SMGs are given in Table~\ref{tab-background}.

\capstartfalse
\begin{deluxetable}{lllll}
\tablecolumns{5}
\tablewidth{245pt}
\tablecaption{Candidate sub-mm galaxies \label{tab-background}}
\tablehead{
   \colhead{Name}
   & \colhead{RA\textsuperscript{a}}
   & \colhead{Dec\textsuperscript{a}}
   & \colhead{$F_{850}$\textsuperscript{b}}
   & \colhead{$\sigma_{850}$\textsuperscript{b}}}
\startdata
SMG J053450.5--100120&  83.7106  &  10.0222 & 13.23 & 2.63 \\
SMG J053552.0--100304&  83.9667  &  10.0510 & 13.57 & 2.84 \\
SMG J053550.4--100443&  83.9600 &   10.0787 & 13.30 & 2.61
\enddata
\tablenotetext{a}{Units of deg (J2000).}                                     
\tablenotetext{b}{Units of mJy beam$^{-1}$.}                                     
\end{deluxetable}
\capstartfalse

\section{DISCUSSION\label{sec-discussion}}

\subsection{Dust Masses\label{sec-dust}}

We can estimate the dust mass of each disk using the 850-$\mu$m continuum fluxes in Table~\ref{tab-flux}. Assuming the dust emission at sub-mm wavelengths is optically thin, the sub-mm continuum flux ($F_{\nu}$) is directly related to the dust mass ($M_{\rm dust}$) as in \cite{1983QJRAS..24..267H}:  

\begin{equation}
M_{\rm dust}=\frac{F_{\nu}d^{2}}{\kappa_{\nu}B_{\nu}(T_{\rm dust})},
\label{eqn-mass1}
\end{equation}

where $B_{\nu}$ is the Planck function and we assume a characteristic dust temperature of $T_{\rm dust}=20$ K, which is the median found for Taurus disks \citep{2005ApJ...631.1134A}. The dust grain opacity, $\kappa_{\nu}$, is taken as 10 cm$^{2}$ g$^{-1}$ at 1000 GHz with an opacity power-law index $\beta=1$ \citep{1990AJ.....99..924B}. Using the distance to \lori{} of $d\approx450$ pc \citep{2001AJ....121.2124D}, we can estimate dust masses from $F_{850}$ (mJy) using:

\begin{equation}
M_{\rm dust} \approx 5.8 \times 10^{-6} F_{850}~{\rm M}_{\odot}.
\label{eqn-mass2}
\end{equation}

This gives a dust mass for the disk around \hd{} of $\sim150$ M$_{\oplus}$ and a 3$\sigma$ upper limit on the average dust mass for all other \lori{} disks of $\sim3$ M$_{\oplus}$.

\subsection{Comparison with other star-forming regions \label{sec-compare}}

\capstartfalse
\begin{figure}
\begin{centering}
\includegraphics[width=8.5cm]{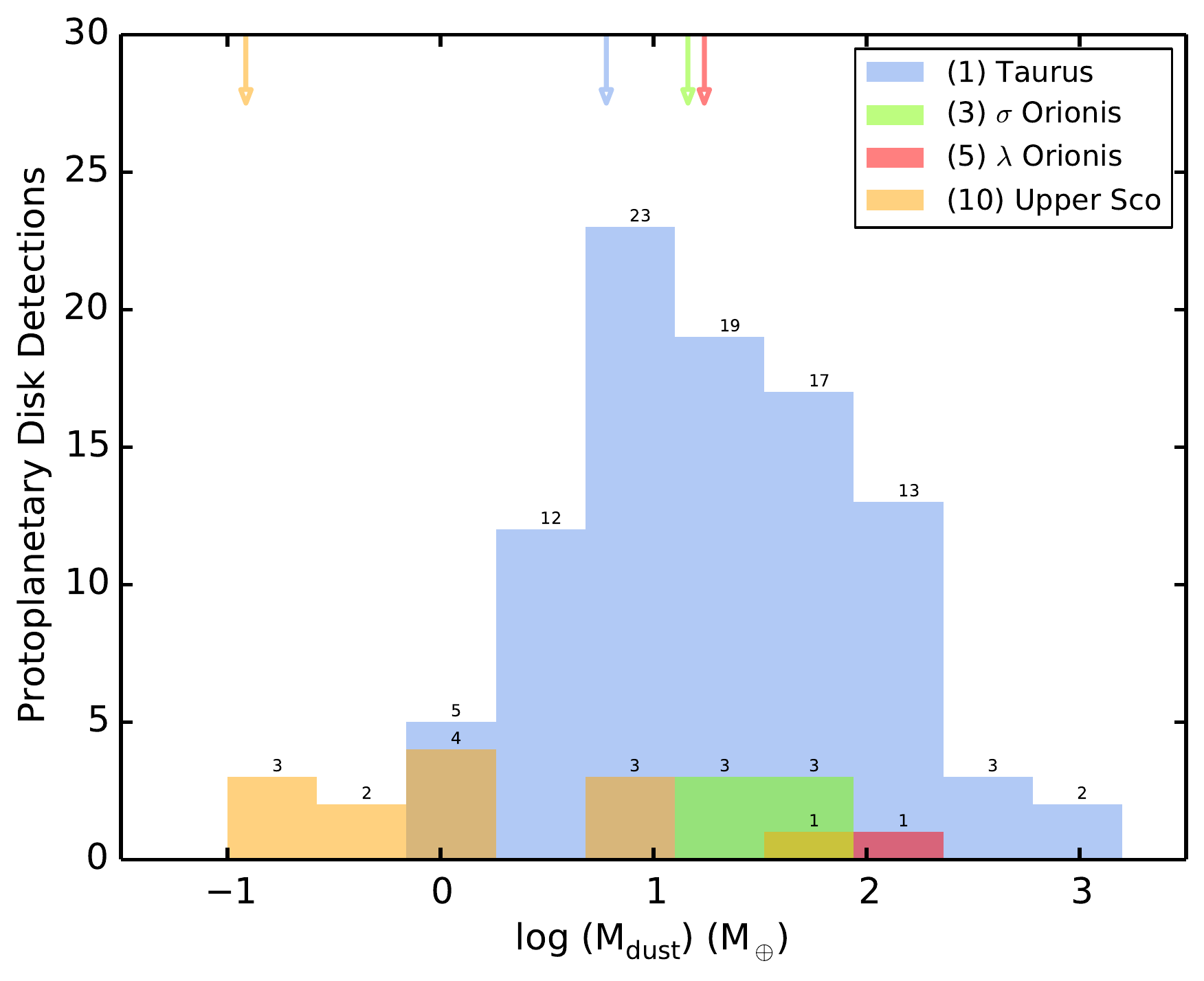}
\caption{\small Dust mass distributions for Class II disks in clusters with ages spanning $\sim1$--10 Myr: Taurus \citep{2005ApJ...631.1134A}, $\sigma$ Orionis \citep{2013MNRAS.435.1671W}, \lori{} (this work), and Upper Sco \citep{2014ApJ...787...42C}. The arrows indicate 3$\sigma$ survey sensitivities and the cluster ages (in Myr) are shown next to the cluster name.\vspace{1.mm}}
\label{fig-masses}
\end{centering}
\end{figure}
\capstartfalse

In Figure~\ref{fig-masses}, we compare the dust mass distribution of optically thick Class II disks in \lori{} with those of several other star-forming regions taken from the literature with ages spanning $\sim1$--10 Myr. We compare only Class II disks as they can still exhibit relatively strong sub-mm emission but lack confusion from large circumstellar envelopes dominating Class 0/I disks. We compare only dust masses, rather than total disk masses, as gas-to-dust ratios are highly uncertain and may vary significantly from disk to disk, even within the same cluster \cite[e.g.,][]{2014ApJ...788...59W}. As slightly different methods are often applied to derive dust masses, we did not take the values directly from the literature. Rather, we translated the published sub-mm continuum fluxes into dust masses using versions of Equation~\ref{eqn-mass1} scaled to the distances of the clusters and the observation wavelengths of the surveys. The survey sensitivities (shown by arrows in Figure~\ref{fig-masses}) are determined from their respective median 3$\sigma$ statistical errors.

The surveys in Figure~\ref{fig-masses} are all sensitive to dust masses $\gtrsim30$ M$_{\oplus}$. This allows us to study how the planet-forming capacity of disks evolves with time, assuming gas-to-dust ratios of $\sim2$--40 \cite[e.g.,][]{2014ApJ...788...59W}. From Figure~\ref{fig-masses}, it appears that over $\sim1$--10 Myr timescales there is substantial evolution in the frequency of disks capable of forming systems with multiple giant planets: such massive disks drop in frequency by roughly an order of magnitude by $\sim5$ Myr, presumably because the disk material has mostly dispersed or is locked in pebbles/planetesimals $>1$ mm in size. During $\sim5$--10 Myr, only outlier disks maintain such high masses. Note that a proper statistical comparison between disk populations should account for different stellar mass distributions, individual disk sensitivities, and survey incompletenesses \citep[e.g.,][]{2013ApJ...771..129A}; however, this approach was impractical here due to the single detection in \lori{}.

Figure~\ref{fig-masses} also highlights the particularly high dust mass of \hd{} for its age, making it an outlier in the \lori{} cluster---and perhaps the disk population in general. Indeed, \hd{} resides far from our survey sensitivity limit and we did not detect any other disks in this region. Moreover, \cite{2009ApJ...707..705H} noted that the 24-$\mu$m excess from \hd{} was over twice as large as other disks around intermediate-mass stars with similar ages. \cite{2015arXiv150200631R} also showed that protoplanetary disks represent only 2\% of the disk population at $\sim5$ Myr for stars $>2$ M$_{\odot}$. Below we discuss possible physical reasons for this outlier status, namely cluster non-membership and gap-opening planet formation.

\subsection{HD 245185\label{sec-det}}

The only significant detection in our 850-$\mu$m survey of \lori{} was \hd{}, a well-known Herbig Ae star of $\sim2.5$ M$_{\odot}$ \citep[e.g.,][]{2012MNRAS.422.2072F}. With a detection significance of $\sim28\sigma$, its dust mass is $\gtrsim50$ times larger than the average \lori{} disk. Its strong sub-mm continuum emission, in addition to its large {\it Spitzer} IR excesses \citep{2009ApJ...707..705H}, suggests that \hd{} hosts a massive primordial disk. However, protoplanetary disks around intermediate-mass stars are supposed to disperse twice as fast compared to low-mass stars \citep{2015arXiv150200631R}, raising the question of why such a massive primordial disk has persisted around \hd{}. In this section we present a more detailed analysis of \hd{} to investigate this question.

\subsubsection{Spectral energy distribution\label{sec-sed}}

\capstartfalse
\begin{figure}
\begin{centering}
\includegraphics[width=8.5cm]{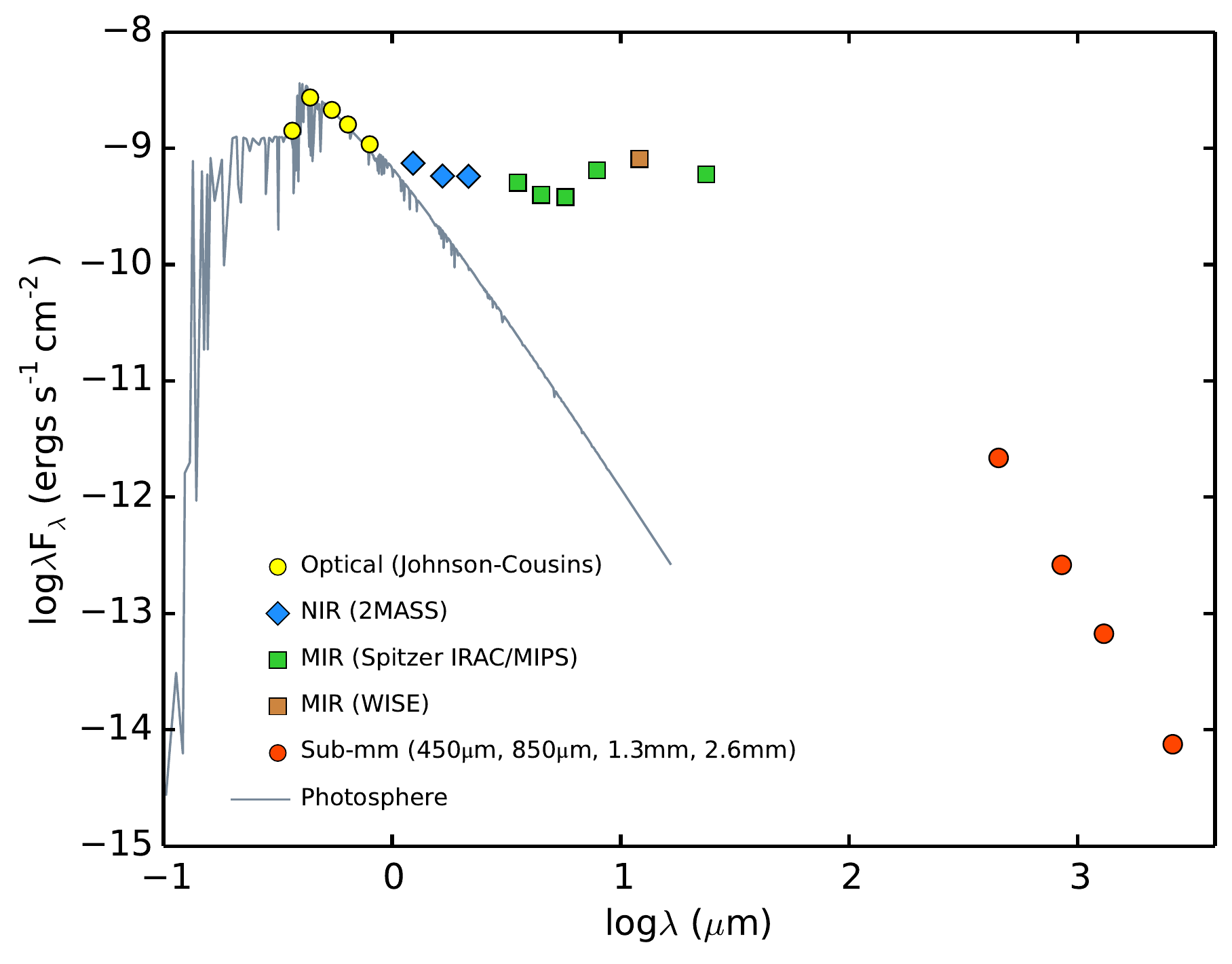}
\caption{\small The optical--mm wavelength SED for \hd{}. Errors are all smaller than the symbol sizes. The gray line is a NextGen model photosphere \citep{1999ApJ...512..377H} for a 9600 K star with log $g = 4.0$ and [M/H] = 0 \citep{2012MNRAS.422.2072F}, scaled to a distance of 450 pc. Extinction corrections are not included due to the low reddening toward \lori{}. \vspace{0.5mm}}
\label{fig-sed}
\end{centering}
\end{figure}
\capstartfalse

Figure~\ref{fig-sed} shows the optical--mm wavelength SED of \hd{} (values given in Table~\ref{tab-sed}). The circumstellar disk around the central star produces the IR excesses above the expected stellar photosphere emission (gray line in Figure~\ref{fig-sed}). \cite{2009ApJ...707..705H} classified the disk as primordial due to its large mid-IR excesses in all {\it Spitzer} bands. Our SED shows that this excess exists even at near-IR wavelengths, which suggests that the disk extends down to very small radii ($\lesssim1$ AU). 

However, the mid-IR dip in the SED is a strong indicator of a pre-transition disk, which has evolved past the primordial disk phase by developing a gap in the disk structure \citep{2014prpl.conf..497E}. The large disk mass ($\sim150$ M$_{\oplus}$; Section~\ref{sec-dust}) rules out a highly evolved debris disk as these typically have much lower dust masses of $\sim10^{-1}$ M$_{\oplus}$ \citep[see Figure 3 in][]{2008ARA&A..46..339W}. We also fit the tail end of the SED to calculate the sub-mm spectral index, assuming $F_{\nu}\propto\nu^{\alpha}$ at these longer wavelengths. We found a shallow slope of $\alpha=2.23\pm0.04$, close to the blackbody value suggesting grain growth to at least mm-size objects \citep{2006ApJ...636.1114D}. This $\alpha$ value is similar to those found for other disks \citep[e.g.,][]{2014prpl.conf..339T}.

Note that we show only one WISE band in Figure~\ref{fig-sed} for clarity, as the others overlap those of {\it Spitzer}. In Table~\ref{tab-sed} we include all four WISE measurements, which are consistent with the shape of the SED traced by the {\it Spitzer} measurements in Figure~\ref{fig-sed}. Thus there is no evidence of \hd{} exhibiting mid-IR variability over a several-yr timescale, which could indicate major collisions occurring in the disk \citep{2015arXiv150305610M}

\capstartfalse
\begin{deluxetable}{lll}
\tablecolumns{3}
\tablewidth{245pt}
\tablecaption{Spectral energy distribution \label{tab-sed}}
\tablehead{
     \colhead{$\lambda_{eff}$}
   & \colhead{$F$}
   & \colhead{Source [Filter]} \\
      \colhead{(\AA$\times10^3$)}
   & \colhead{(Jy)}
   & \colhead{} }
\startdata
3.660  &  $0.1725  \pm  0.0048$  & \text{\citealt{2001A&A...380..609D}} [$U$]  \\
4.380  &  $0.3989  \pm  0.0110$  &  \text{\citealt{2001A&A...380..609D}} [$B$] \\
5.450  &  $0.3878  \pm  0.0107$  &  \text{\citealt{2001A&A...380..609D}} [$V$] \\
6.410  &  $0.3422  \pm  0.0094$  &  \text{\citealt{2001A&A...380..609D}} [$R_{c}$] \\
7.980 &  $0.2878  \pm  0.0080$  &   \text{\citealt{2001A&A...380..609D}} [$I_{c}$]  \\
12.35 &  $0.3063  \pm  0.0085$  &  2MASS [$J$] \\
16.62 &  $0.3197  \pm  0.0147$  &  2MASS [$H$]  \\
21.59 &  $0.4130  \pm  0.0152$  &  2MASS [$K_{\rm S}$]  \\
33.68 &  $0.5414  \pm  0.0299$  &  WISE [1]  \\
35.50 &  $0.5983  \pm  0.0165$  &  {\it Spitzer} [IRAC-1]  \\
44.93 &  $0.5945  \pm  0.0164$  &  {\it Spitzer} [IRAC-2]  \\
46.18 &  $0.6073  \pm  0.0145$  &  WISE [2]  \\
57.31 &  $0.7269  \pm  0.0201$  &  {\it Spitzer} [IRAC-3]  \\
78.72 &  $1.6945  \pm  0.0468$  &  {\it Spitzer} [IRAC-4]  \\
120.8 &  $3.2591  \pm  0.0360$  &  WISE [3]  \\
221.9 &  $5.2134  \pm  0.0720$  &  WISE [4]  \\
236.8 &  $4.7021  \pm  0.0001$  &  {\it Spitzer} [MIPS 24 $\mu$m] \\
4500 &  $0.3262  \pm  0.0524$  &  This work \\
8500 &  $0.0741  \pm  0.0042$  &  This work  \\
13000 &  $0.0290  \pm  0.0058$  &  This work  \\
26000 &  $0.0065  \pm  0.0012$  &  \text{\citealt{1997ApJ...490..792M}} \\
\enddata
\end{deluxetable}
\capstartfalse

\subsubsection{CO emission\label{sec-gas2dust}}

We detected weak CO 2--1 line emission from \hd{} (Section~\ref{sec-smaresults}), indicating the presence of molecular gas in the disk. Converting integrated CO 2--1 line flux into a total gas mass is complicated by the optical thickness of the line emission. Instead, our best constraints come from comparing our upper limits on the CO isotopologues to the grid of models in \cite{2014ApJ...788...59W}, who used these optically thin lines to predict total gas masses to within a factor of three. Our $3\sigma$ upper limits to the $^{13}$CO and C$^{18}$O line luminosities for \hd{} ($\sim8\times10^4$ Jy km s$^{-1}$ pc$^2$) only cover model disks with $M_{\rm gas} <1$ M$_{\rm Jup}$ (see Figure 6 in \citealt{2014ApJ...788...59W}). This implies a total gas mass for \hd{} of $\lesssim1$ M$_{\rm Jup}$, which when combined with our measured dust mass of $\sim150$ M$_{\oplus}$ (Section~\ref{sec-dust}) gives a notably low gas-to-dust ratio of $\lesssim2$. Although we cannot rule out a small, optically thick disk as an alternative explanation for the low CO 2--1 line emission (we only have an upper limit on the disk radius of $\sim200$ AU; Section~\ref{sec-smaresults}), disks around Herbig Ae stars are typically $>100$ AU in radius. 

This substantial gas depletion sets \hd{} apart from other disks around Herbig Ae stars of similar age, which have been found to remain relatively gas rich. For example, the transition disk HD 169142 ($\sim6$ Myr) has $M_{\rm gas}\sim20$ M$_{\rm Jup}$  \citep{2008A&A...491..219P} and $M_{\rm dust}\sim20$ M$_{\rm Jup}$ \citep{2006AJ....131.2290R}, thereby maintaining its initial ISM gas-to-dust ratio of $\sim100$. The massive disk HD 163296 ($\sim4$ Myr) has $M_{\rm gas}\sim50$ M$_{\rm Jup}$ and $M_{\rm dust}\sim0.3$ M$_{\rm Jup}$ \citep{2014ApJ...788...59W}, suggesting a gas-to-dust ratio of $\sim170$ (\citealt{2011ApJ...740...84Q} also estimate a ratio of $\sim150$). The distinction from \hd{} is significant, even when considering that gas-to-dust ratios are typically uncertain by a factor of $\sim10$. 

\capstartfalse
\begin{figure}
\begin{centering}
\includegraphics[width=8.5cm]{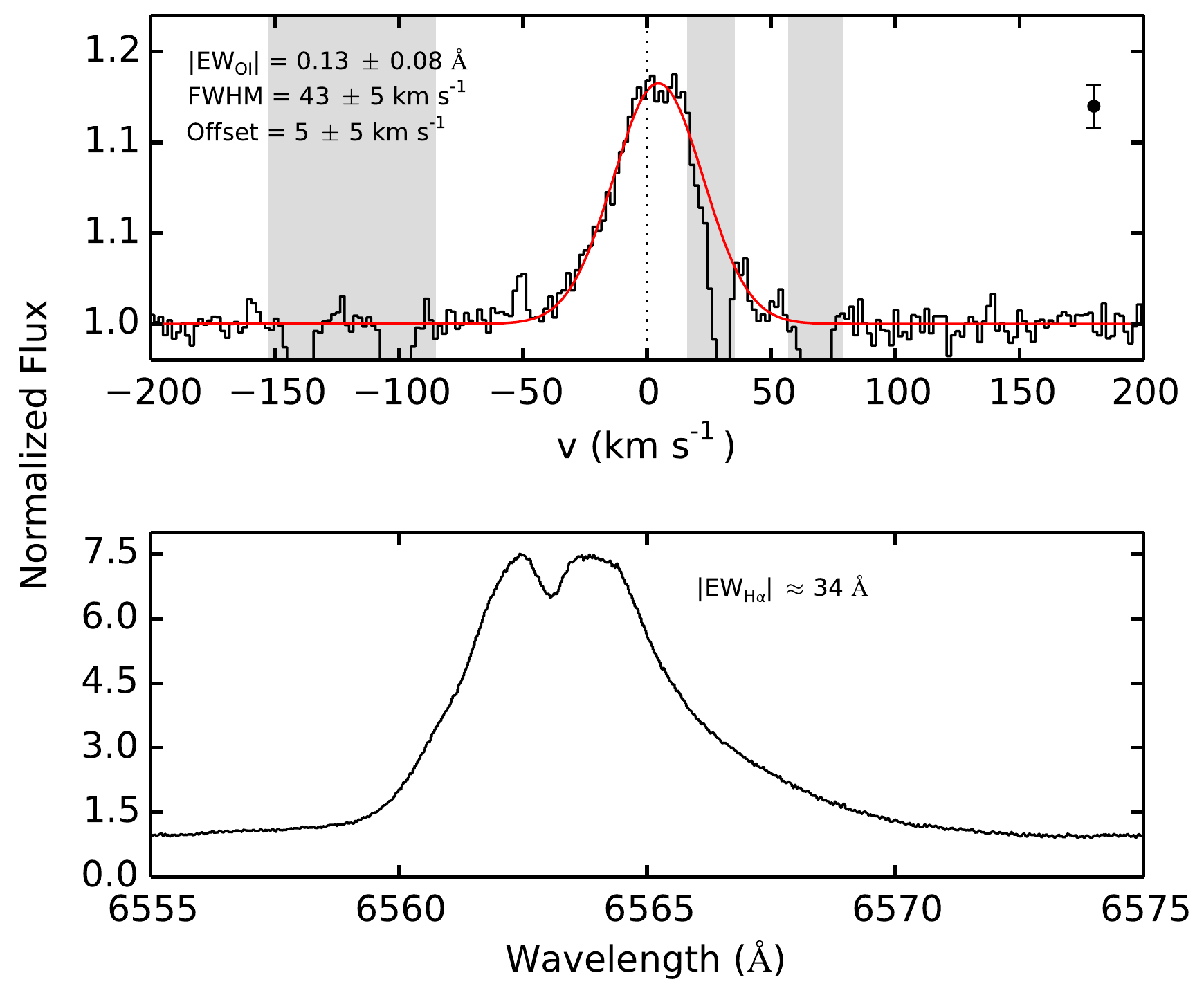}
\caption{\small Archival CFHT/ESPaDOnS spectra of \hd{}. Top panel shows \OI{} 6300{\AA} line emission with a median error bar given for reference. The $x$-axis is centered on the stellar radial velocity of $v_{\rm rad}=11.0$ km s$^{-1}$ (Section~\ref{sec-smaresults}); the peak of the Gaussian fit (red line) after masking out telluric absorption features (gray regions) is offset by 5 km s$^{-1}$ but within measurement errors. Bottom panel shows H$\alpha$ 6563{\AA} line emission, likely from accretion as it is strong, broadened, double-peaked, and axisymmetric.$\vspace{0.8 mm}$}
\label{fig-oxygen}
\end{centering}
\end{figure}
\capstartfalse

\subsubsection{ \OI{} \& H$\alpha$ emission \label{sec-OI}}

To further investigate the gas component of the disk around \hd{}, we analyzed the non-thermal \OI{} 6300{\AA} emission line. This forbidden emission arises when UV photons from the host star photodissociate OH molecules in the surface layers of the circumstellar disk to create excited oxygen atoms, some of which are in the upper state of the 6300{\AA} line \citep{1984ApJ...277..576V}. \OI{} 6300{\AA} line luminosities have been empirically correlated to accretion rates for Herbig Ae stars \citep{2011A&A...535A..99M}, which can be physically explained if the UV excess from the accretion shock is illuminating the disk to excite the oxygen atoms. Although \OI{} 6300{\AA} emission is clearly not a good measure of total gas mass, it can be used to investigate the presence of gas in the inner disk of \hd{} despite its depleted overall gas content (Section~\ref{sec-gas2dust}).

\cite{2005A&A...436..209A} previously analyzed the \OI{} 6300{\AA} emission line from \hd{} using a high-resolution spectrum ($R=45,000$) but with limited S/N (see their Figure 2). We therefore re-visited this analysis with higher resolution ($R=65,000$) and higher S/N ($\sim80$ per resolution element at 6300 {\AA}) archival spectra from the Echelle Spectropolarimetric Device for the Observation of Stars (ESPaDOnS) instrument on the Canada-France-Hawaii Telescope (CFHT). We median combined the four spectra taken on 2005-02-20, normalized the spectrum to the fitted continuum, masked out telluric absorption features, and fit a Gaussian to the spectral line. The resulting spectrum and fit are shown in Figure~\ref{fig-oxygen}. We found $\lvert {\rm EW} \rvert =0.13\pm0.08$ {\AA} and ${\rm FWHM}=43\pm5$ km s$^{-1}$, both of which are typical for Herbig Ae/Be stars with pure emission line profiles \citep[e.g., group I in][]{2005A&A...436..209A}. The offset of the emission line peak from our derived stellar radial velocity of $v_{\rm rad}=11.0$ km s$^{-1}$ (Section~\ref{sec-smaresults}) is within measurement errors.

These results suggest that HD 245185 has the typical \OI{} 6300{\AA} line emission expected from disks around accreting Herbig Ae stars. This accretion signature is supported by the H$\alpha$ 6563{\AA} line emission from \hd{}, also seen in its archival CFHT/ESPaDOnS spectrum (Figure~\ref{fig-oxygen}). The H$\alpha$ emission line is strong ($\lvert {\rm EW}_{\rm{H} \alpha} \rvert \approx 34$ {\AA}), broadened, double-peaked, and asymmetric. Indeed, \cite{2006ApJ...653..657M} previously found $\lvert {\rm EW}_{\rm{H} \alpha} \rvert \approx 25$ {\AA} and suggested this as a signature of accretion due to correlations between H$\alpha$ emission and near-IR excess. \cite{2011AJ....141...46D} also used veiling of the Balmer discontinuity to derive an accretion rate of $0.63\pm0.41\times10^{-7}$ M$_{\odot}$ yr$^{-1}$, which is typical for \hd{}'s stellar mass and age \cite[e.g.,][]{2011A&A...535A..99M,2011ApJ...734...22L}. This ongoing accretion is surprising given the depleted overall gas content for this disk (Section~\ref{sec-gas2dust}), and suggests that we may be catching \hd{} at the final stages of disk clearing.

\subsubsection{Cluster membership \label{sec-membership}}

\hd{} is a clear outlier in the \lori{} star-forming region in terms of its circumstellar disk properties. This brings into question whether \hd{} is a true member of the cluster. Dynamically, \hd{} is a debatable member of \lori{}. The proper motion of \hd{} is $\mu_{\alpha}cos(\delta)\approx0.67\pm1.3$ and $\mu_{\delta}\approx-1.60\pm1.3$ mas yr$^{-1}$ \citep{2008A&A...488..401R}, which is in very good agreement with the average proper motion of the early-type members of \lori{} \cite[see Figure 1 in][]{2009ApJ...707..705H}. However, the precise radial velocity derived from our SMA observations of $v_{\rm rad}=11\pm0.1$ km s$^{-1}$ (Section~\ref{sec-smaresults}) is significantly different from the tight distribution of radial velocities for \lori{} cluster members of $\overline{v}_{\rm rad}=24.3\pm2.3$ km s$^{-1}$ \citep{2001AJ....121.2124D}. Peculiar kinematics of high-mass stars in the \lori{} cluster have been previously attributed to a supernova that may have occurred $\sim1$ Myr ago in the \lori{} stellar system \citep[e.g.,][]{2001AJ....121.2124D,1996A&A...309..892C}, suggesting kinematics alone are an unreliable criteria for membership. However, \hd{} is unlikely to have been affected by such a supernova, at least kinematically, as the projected distance between \hd{} and the \lori{} star is almost a parsec. 

Interestingly, the kinematics of \hd{} are consistent with those of the nearby, younger Taurus region. Taurus has $\overline{v}_{\rm rad}=16\pm6$ km s$^{-1}$ \citep{2006A&A...460..499B}, which is consistent with the $v_{\rm rad}$ of \hd{} derived in this work. The proper motion of \hd{} is also within $3\sigma$ of the average proper motion of Taurus members \citep{2006A&A...460..499B}. Indeed, \cite{2001AJ....121.2124D} found in their kinematic study of \lori{} a grouping of candidate members with strong Li absorption and $v_{\rm rad}$ values similar to that of Taurus, although with random spatial distribution. However, Taurus is much closer than \lori{} at a distance of $\sim140$ pc. Moving \hd{} $\sim3\times$ closer would prescribe a stellar luminosity several factors below the main sequence for the well-determined A0/A1 spectral type of \hd{} \cite[e.g.,][]{2006ApJ...653..657M,2012MNRAS.422.2072F} or require $\sim2.5$ magnitudes of extinction.

Age estimates for \hd{} are consistent with the $\sim5$ Myr age of \lori{}, but also within the range of ages seen in Taurus \citep{2013ApJ...771..129A}. Several authors have estimated the age of \hd{}, for example: $6.9\pm2.5$ \citep{2013MNRAS.429.1001A} and $5.5\pm2.0$ \citep{2012MNRAS.422.2072F}. Moreover, \cite{2011A&A...530A.150F} showed that \hd{} sits on the \lori{} cluster sequence in optical and IR color-magnitude diagrams and color-color diagrams (see their Figure 2). 

Based on the above information, we cannot definitively say whether \hd{} is a true member of the \lori{} cluster, or if it is an interloper from a nearby star-forming region such as Taurus.

\subsection{Reconciling \hd{} with planet formation? \label{sec-debris}}

Disk simulations predict that forming planets clear their orbits \cite[e.g.,][]{2011ApJ...729...47Z} producing SEDs with mid-IR dips corresponding to resolved cavities in sub-mm/mm images \citep{2011ApJ...732...42A}; this is observationally supported by protoplanetary candidates imaged inside transition disk cavities \cite[e.g.,][]{2012ApJ...745....5K}. Dust is thought to be filtered through the cavities, such that large grains ($\gtrsim$ sub-mm) are built up at the outer cavity edges while smaller particles and gas are allowed to flow across \cite[e.g.,][]{2006MNRAS.373.1619R}. 

These signatures of disk clearing from planet formation may explain our observations of \hd{}. The mid-IR dip in the SED (Figure~\ref{fig-sed}) would originate from planet(s) clearing cavities in the disk. The high observed dust mass (Section~\ref{sec-dust}) would result from sub-mm/mm grains containing the bulk of the dust mass being trapped in the outer disk, where they are most effectively probed by sub-mm surveys. The gas depletion (Section~\ref{sec-gas2dust}) yet ongoing accretion (Section~\ref{sec-OI}) would be from dust filtration allowing gas (and small particles) to continue flowing onto the star.

\section{SUMMARY \& FUTURE WORK\label{sec-conclusions}}

We presented results from an 850-$\mu$m survey of the $\sim5$ Myr old \lori{} cluster using JCMT/SCUBA-2. Our $\sim0\fdg5$-diameter survey region, with a mean rms noise level of $\sim2.9$ mJy beam$^{-1}$, contained 36 (out of 59) cluster members with IR excesses indicative of circumstellar disks. We found only one significant detection, the Herbig Ae star \hd{}, with an inferred dust mass of $\sim150$ M$_{\oplus}$. The $3\sigma$ upper limit on the average dust mass of the individually non-detected sources was only $\sim3$ M$_{\oplus}$. Thus, despite $\sim15$ \% of \lori{} cluster members bearing circumstellar disks identified via their IR excesses, the average disk at $\sim5$ Myr no longer stores sufficient dust and gas reservoirs to form giant planets and perhaps even super Earths. Comparing the dust mass distribution of disks in \lori{} to those of other star-forming regions with ages spanning $\sim1$--10 Myr showed that the frequency of disks capable of forming systems with multiple giant planets drops by roughly an order of magnitude by $\sim5$ Myr.

Our survey showed that \hd{} is a clear outlier in the \lori{} cluster, as its sub-mm emission was far above our survey sensitivity and we did not detect any other disks in the region. Our follow-up SMA observations of \hd{} provided upper limits on optically thin CO isotopologue lines luminosities, which we used to infer a gas mass of $\lesssim1$ M$_{\rm Jup}$. This implies an exceptionally low gas-to-dust ratio of $\lesssim2$, which is $\gtrsim50\times$ lower than the presumed initial ISM value. This overall gas depletion sets \hd{} apart from other Herbig Ae disks of similar age, which have been found to be gas rich. Indeed, although \hd{}'s strong sub-mm continuum emission and large {\it Spitzer} IR excesses pointed to a massive primordial disk, the mid-IR dip in its SED as well as its exceptionally low gas-to-dust ratio are clear indicators of a more evolved disk. However, its dust mass is still orders of magnitude higher than that of debris disks. Its optical emission lines indicative of ongoing accretion, despite significant gas depletion, suggest we may be catching \hd{} during the final stages of disk clearing. One possible explanation for our observations is that the disk around \hd{} is forming gap-opening planets.

The higher sensitivity of the Atacama Large Millimeter Array (ALMA) could possibly determine the residual amounts of gas and dust in these older disks and thus place stronger constraints on planet-formation timescales. ALMA may also spatially resolve the \hd{} disk, which could reveal signatures of ongoing planet formation hypothesized in this work.

\begin{acknowledgements}
This work is supported by NSF and NASA through grants AST-1208911 and NNX15AC92G, respectively. The James Clerk Maxwell Telescope is operated by the Joint Astronomy Centre on behalf of the Science and Technology Facilities Council of the United Kingdom, the Netherlands Organization for Scientific Research, and the National Research Council of Canada. Additional funds for the construction of SCUBA-2 were provided by the Canada Foundation for Innovation. The Submillimeter Array is a joint project between the Smithsonian Astrophysical Observatory and the Academia Sinica Institute of Astronomy and Astrophysics and is funded by the Smithsonian Institution and the Academia Sinica. This research utilized the NASA Astrophysics Data System, SIMBAD database, and Vizier catalogue access tool.
\end{acknowledgements}

\bibliography{ms.bib}

\end{document}